\documentclass[aps,balance,twocolumn,pra,tightenlines,floatfix,showpacs]{revtex4-1}
\usepackage[margin=0.75in,footskip=0.2in]{geometry}
\usepackage{graphicx}
\usepackage[english]{babel}
\usepackage{amsmath,amssymb,float}
\usepackage{slashed}
\usepackage{pdfpages}
\usepackage{hyperref}
\usepackage{dcolumn}
\usepackage[caption=false]{subfig}
\hypersetup
  {citecolor = {blue},
  colorlinks = {true}, 
  urlcolor = {red}}
\def\p{\partial}
\def\mb{\mathbf}
\makeatletter
\AtBeginDocument{\let\LS@rot\@undefined}
\makeatother

\begin{document}
\title{Exact correlation functions in the cuprate pseudogap phase:\\
combined effects of charge order and pairing}

\author{Rufus Boyack, Chien-Te Wu, Peter Scherpelz, and K. Levin}
\affiliation{$^1$James Franck Institute, University of Chicago, Chicago, Illinois 60637, USA}

\begin{abstract}
There is a multiplicity of charge ordered, pairing-based 
or pair density wave theories of the cuprate pseudogap,
albeit arising from different microscopic mechanisms.
For mean field schemes (of which there are many)
we demonstrate here that they have precise implications for
two body physics in the same way that they are able
to address the one body physics of photoemission spectroscopy.
This follows because the full vertex can be obtained exactly from 
the Ward-Takahashi identity.
As an illustration, we present the spin response functions, finding
that a recently proposed pair density wave (Amperean pairing) 
scheme is readily distinguishable
from other related scenarios. 
\end{abstract}
\maketitle

\textit{Introduction.$-$} 
A number of theories associated with the 
cuprate pseudogap phase have recently been suggested, based on 
now widely observed charge order \cite{Sawatzky13,Hudson13,Comin,Keimer}.
While the underlying physics
may be different, what emerges rather generally
are BCS-based pairing theories of the normal state
with band-structure reconstruction \cite{SCZhang,Lee_2014,YRZ}.
Distinguishing between theories has mostly been
based on angle resolved photoemission spectroscopy 
(ARPES) \cite{arpesstanford}.
However, the majority of data available
for the cuprates involves two particle properties: for
example, the optical absorption \cite{Corson1999},
diamagnetism \cite{Ong2}, quasi-particle interference in STM 
\cite{Davis}, neutron \cite{Birg,Dai,Keimer}
and inelastic x-ray scattering in the charge \cite{Comin}
and spin \cite{Keimerparamag} sectors.

In this paper we use the Ward-Takahashi identity (WTI) \cite{Ryder,Schrieffersbook} to
develop precise two body response functions for these pairing based pseudogap theories.
Such exact response functions make it possible to address two particle cuprate experiments,
including the list above, from the perspective of many different theories. 
As an illustration, we compute the spin-spin correlation functions relevant to neutron
scattering in three pseudogap scenarios.
That the response functions analytically satisfy the $f$-sum rule 
provides the confidence that there are no
missing Feynman diagrams or significant numerical inaccuracies.

By comparing the Amperean pairing scheme \cite{Lee_2014},
and that of Yang, Rice and Zhang \cite{YRZ} with a simple $d$-wave
pseudogap scenario, we find that the Amperean theory leads to a relatively
featureless neutron cross section in contrast to the peaks (at and near the antiferromagnetic wave vector),
found for the other two theories.

In this Amperean pairing scheme \cite{Lee_2014} the mean field self energy is 
\begin{widetext}\begin{align}\label{eq:LSE}
\Sigma_{pg}(K)&=\frac{\Delta_{1}^2}{\omega+\xi_{\mb{k-p}}-\frac{\Delta_{2}^2}{\omega-\xi_{\mb{k}-2\mb{p}}}}
+\frac{\Delta_{2}^2}{\omega+\xi_{\mb{k+p}}-\frac{\Delta_{1}^2}{\omega-\xi_{\mb{k}+2\mb{p}}}}+
\frac{C_{1}^2}{\omega-\xi_{\mb{k}+2\mb{p}}-\frac{\Delta_{1}^2}{\omega+\xi_{\mb{k+p}}}}
+\frac{C_{2}^2}{\omega-\xi_{\mb{k}-2\mb{p}}-\frac{\Delta_{2}^2}{\omega+\xi_{\mb{k-p}}}}\nonumber\\
&+\frac{2\Delta_{1}\Delta_{2}C_{1}}{\left(\omega+\xi_{\mb{k+p}}\right)\left(\omega-\xi_{\mb{k+2p}}\right)-\Delta_{1}^2}
+\frac{2\Delta_{1}\Delta_{2}C_{2}}{\left(\omega+\xi_{\mb{k-p}}\right)\left(\omega-\xi_{\mb{k-2p}}\right)-\Delta_{2}^2}.
\end{align}
\end{widetext}

We single this particular theory out as an example which is
complex and therefore somewhat more inclusive. 
In Eq.~(\ref{eq:LSE}) $\Sigma_{pg}(K)$ is expressed in terms of two different finite momentum ($\mb{p}$) pseudogaps,
$\Delta_1 \equiv \Delta_{\mb{p}}$ and $\Delta_2 \equiv \Delta_{-\mb{p}}$.
In addition we have introduced charge density wave (CDW) amplitudes $C_{1} \equiv C_{2\mb{p}}$ and $C_{2} \equiv C_{-2\mb{p}}$. 
From the self energy, the full (inverse)
Green's function can be deduced: $G^{-1}(K)\equiv G_{0}^{-1}(K)-\Sigma_{pg}(K)
=\omega - \xi_\mb{k} - \Sigma_{pg}(K)$.
This then determines the renormalized band-structure, which can be compared with ARPES experiments.
One can similarly add other mean field contributions such as that
related to an SDW \cite{Halperin_2003} or even a DDW \cite{Laughlin}.

It is observed from Eq.~(\ref{eq:LSE})
that in the Amperean pairing case a BCS-like self energy appears in
a continued fraction form within the self energy itself.
There are analogies with the approach of 
Yang, Rice and Zhang (YRZ) \cite{YRZ} 
in the limit that only one gap term is present, say $\Delta_1$,
and when the CDW ordering is absent. 
Importantly, this single gap self energy
involves two types of dispersion relations, 
so that the pairing term leads to pockets or
a reconstruction of the band-structure.
For a simpler $d$-wave pseudogap, with a single gap model, 
both of these dispersion relations are taken to be the same, as was studied
microscopically \cite{ourreview} and phenomenologically \cite{Normanphenom}.
A central contribution of this paper is to show how, via
two particle properties, important distinctions
between these three different pseudogap theories can be established.

While it is argued to be appropriate for the pseudogap phase \cite{Lee_2014},
the self energy of Eq. (\ref{eq:LSE}) is indistinguishable from that of a superconducting state.
It is important, then, to ensure that this form for $\Sigma_{pg}$ 
does not correspond to an ordered phase.
Phase fluctuations have been phenomenologically
invoked \cite{SCZhang,Lee_2014} to destroy order.
Regardless of this phenomenology there is a quantitative
constraint to be satisfied: the absence
of a Meissner effect above $T_{C}$
implies that the zero frequency and zero momentum 
current-current correlation function satisfies
$-\tensor{P}(0)=\left(\tensor{\frac{n}{m}}\right)_{\rm dia}
\equiv2\sum_K\frac{\partial^2\xi_{\mb{k}}}{\partial{\mb{k}}\partial{\mb{k}}}G(K)$, so that there is a precise cancellation 
between the diamagnetic and paramagnetic current-current
correlation functions in the normal state.

Performing integration by parts \cite{discuss}
and using the identity $\partial G(K)/\partial\mb{k}=-G^2(K)\partial 
G^{-1}(K)/\partial\mb{k}$ 
then yields the following expression for $\tensor{P}(0)$:
\begin{equation}\label{eq:RF0}\tensor{P}(0)
=2\sum_{K}G^{2}(K)\biggl\{\frac{\p\xi_{\mb{k}}}{\p\mb{k}}+
\frac{\p\Sigma_{pg}(K)}{\p\mb{k}}\biggr\}\frac{\p\xi_{\mb{k}}}{\p\mb{k}}.
\end{equation} 
Here $K=(\omega,\mb{k})$. Given the self energy from Eq. (\ref{eq:LSE}),
it is then straightforward to arrive at the quantity $\tensor{P}(0)$:
\begin{align}\label{eq:RF02}
&\tensor{P}(0)=2\sum_{K}G^{2}(K)\biggl\{\frac{\p\xi_{\mb{k}}}{\p\mb{k}}-
\Delta^{2}_{1}G^{2}_{1,1}(-K)\frac{\p\xi_{\mb{k},2}}{\p\mb{k}}\nonumber\\
&+\Delta^{2}_{1}\Delta^{2}_{2}G^{2}_{1,1}(-K)G^{2}_{0,4}(K)\frac{\p\xi_{\mb{k},4}}{\p\mb{k}} 
-\Delta^{2}_{2}G^{2}_{1,2}(-K)\frac{\p\xi_{\mb{k},1}}{\p\mb{k}}\nonumber \\
&+\Delta^{2}_{2}\Delta^{2}_{1}G^{2}_{1,2}(-K)G^{2}_{0,3}(K)\frac{\p\xi_{\mb{k},3}}{\p\mb{k}}
\biggr\}\frac{\p\xi_{\mb{k}}}{\p\mb{k}}.
\end{align}
For simplicity, throughout the main text we set $C_{1}=C_{2}=0$ and present the complete expressions in the Supplemental Material.
Here we have defined the following four bare (inverse) Green's functions
$G^{-1}_{0,i}(K)=(\omega-\xi_{\mb{k},i}),\quad i\in \{ 1,2,3,4 \}$,
where $\xi_{\mb{k},1} = \xi_{\mb{k+p}}$, $\xi_{\mb{k},2}=\xi_{\mb{k-p}}$,
$\xi_{\mb{k},3}=\xi_{\mb{k}+2\mb{p}}$, $\xi_{\mb{k},4}=\xi_{\mb{k}-2\mb{p}}$
are four dispersion relations. (The usual bare inverse Green's function is denoted by
$G^{-1}_{0}(K)=\omega-\xi_{\mb{k}}=\omega-\epsilon_{\mb{k}}+\mu$.)
The partially dressed Green's functions (which are neither bare nor full)
associated with Eq. (\ref{eq:LSE}) are
\begin{align}
G^{-1}_{1,1}(K)&={\omega-\xi_{\mb{k},1}}-\frac{\Delta_{2}^2}{\omega+\xi_{-\mb{k},4}},
\\G^{-1}_{1,2}(K)&={\omega-\xi_{\mb{k},2}}-\frac{\Delta_{1}^2}{\omega+\xi_{-\mb{k},3}}.
\end{align}

In terms of these partially dressed Green's functions 
the self energy in Eq. (\ref{eq:LSE}) for the case where $C_{1}=C_{2}=0$ 
has the compact form $\Sigma_{pg}(K)=-\Delta^2_{1}G_{1,1}(-K)-\Delta^2_{2}G_{1,2}(-K)$.
The quantity $\tensor{P}(0)$ in Eq.~(\ref{eq:RF02}) provides
a template for the form of the Feynman diagrams that we will find in $\tensor{P}(Q)$.

\textit{Ward-Takahashi identity (WTI) for the full vertex.$-$}
The exact expression for the current-current correlation function, $\tensor{P}(Q)$, is contained in
the response functions written as
\begin{equation}\label{eq:RF}
P^{\mu\nu}(Q)=2\sum_{K}\Gamma^{\mu}(\widetilde{K},K)G(K)\gamma^{\nu}(K,\widetilde{K})
G(\widetilde{K}).\end{equation}
Throughout the text, we set $\widetilde{K}\equiv K+Q$.
The full vertex in four-vector notation is
given by $\Gamma^{\mu}(\widetilde{K},K)=(\Gamma^{0}(\widetilde{K},K),\pmb\Gamma(\widetilde{K},K))$, where the
first argument denotes the incoming momentum and the second argument, the outgoing momentum.
Here the quantity $\gamma^{\nu}(K,\widetilde{K})$
represents the bare vertex.

The full response kernel is $K^{\mu\nu}(Q)\equiv P^{\mu\nu}(Q)+
\left({\tfrac{n}{m}}\right)^{\mu\nu}_{\rm dia}(1-\delta_{0,\nu}\delta_{\mu,\nu}),$
where there is no summation over indices in the second term.
The Ward-Takahashi identity in quantum field theory is a diagrammatic identity 
that imposes a symmetry between response functions. The particular symmetry we are 
interested in is the $\rm U(1)_{EM}$ abelian gauge symmetry \cite{Ryder}. 
As we shall show, satisfying the WTI also leads to          
manifestly sum rule consistent response functions.               
Charge conservation is an exact relation between the current-current and 
density-density response functions that follows from this 
$\rm U(1)_{EM}$ symmetry. 
The WTI reflects this charge conservation which imposes the constraint:
$\Omega K^{0\nu}+i\mathrm{div}_{\mb{q}}K^{j\nu}=0.$
The WTI, for the vertex $\Gamma^{\mu}(\widetilde{K},K)$, on a lattice is 
\begin{align}&\Omega\Gamma^{0}(\widetilde{K},K)+i\mathrm{div}_{\mb{q}}\pmb\Gamma(\widetilde{K},K)\nonumber\\&=G^{-1}(\widetilde{K})-G^{-1}(K),\nonumber\\
&=\Omega+i\mathrm{div}_{\mb{q}}\pmb\gamma(\widetilde{K},K)+\Sigma_{pg}(K)-\Sigma_{pg}(\widetilde{K}).\end{align}
The WTI for the bare vertex $\gamma^{\mu}(\widetilde{K},K)$ is
$\Omega+i\mathrm{div}_{\mb{q}}\pmb\gamma =G^{-1}_{0}(\widetilde{K})-G^{-1}_{0}(K)
=\Omega-\xi_{\mb{k+q}}+\xi_{\mb{k}}$.
Similarly we introduce the bare vertices
$\gamma^{\mu}_{i}(\widetilde{K},K)$ associated with the dispersion relations $\xi_{\mb{k},i}$.
Here $\mathrm{div}_{\mb{q}}\pmb\Gamma(\widetilde{K},K)$, complicated due to
lattice effects, is the Fourier transform of 
the divergence of $\pmb\Gamma$.

In the limit $Q\rightarrow0$, the Ward-Takahashi identity reduces to the Ward identity: 
$\delta\Gamma^{\mu}(K,K)\equiv\Gamma^{\mu} (K,K) -\gamma^{\mu}(K,K)=-\p\Sigma_{pg}(K)/\p k_{\mu}.$
This is fully consistent with the arguments leading up to
the no-Meissner constraint in Eq. (\ref{eq:RF0}).
In this continuum limit, ($q\rightarrow0$) the WTI and charge conservation have
familiar forms: $q_{\mu}\Gamma^{\mu}(\widetilde{K},K)=G^{-1}(\widetilde{K})-G^{-1}(K)$ and $q_{\mu}K^{\mu\nu}(Q)=0$.

We emphasize that, given
an arbitrary self energy, solving the WTI analytically for the full vertex
$\Gamma^{\mu}(\widetilde{K},K)$ is generally not possible. However, there is
a well-defined procedure to determine this vertex in principle.
One inserts the bare vertex in all possible places in the self energy 
diagram and sums the resulting series of diagrams. 
For the class of theories considered in this paper
$\Sigma$ itself does \textit{not} depend on the full Green's function $G(\Sigma)$,
but rather depends on the bare Green's functions $G_0$ and their simple extensions;
this is associated with generalized mean field theories.
For example, in strict BCS theory $\Sigma(K) = -\Delta^2 G_0(-K)$.

Importantly, it follows that
in the BCS-like theories of interest here, the full vertex, 
$\Gamma^{\mu}(\widetilde{K},K)$, can be deduced from the equivalent
WTI by considering only finitely many loop diagrams. 
We illustrate this procedure specifically for the first term in the Amperean self energy in Eq. (\ref{eq:LSE}).
Using the form of the self energy, along with the bare WTI, we have: $\Sigma_{1}(K)-\Sigma_{1}(\widetilde{K})$
\begin{align}
&=\Delta^{2}_{1}G_{1,1}(-K)G_{1,1}(-\widetilde{K})\{[\Omega+\xi_{\mb{k+q-p}}-\xi_{\mb{k-p}}]\nonumber\\&+\Delta^{2}_{2}G_{0,4}(K)G_{0,4}(\widetilde{K})[\Omega-\xi_{\mb{k+q}-2\mb{p}}+\xi_{\mb{k}-2\mb{p}}]\}\nonumber\\&=\Delta^{2}_{1}G_{1,1}(-K)G_{1,1}(-\widetilde{K})\{
[\Omega+i\mathrm{div}_{\mb{q}}\pmb\gamma_{1}(-K,-\widetilde{K})]\nonumber \\
&+\Delta^{2}_{2}G_{0,4}(K)G_{0,4}(\widetilde{K})[\Omega+i\mathrm{div}_{\mb{q}}\pmb\gamma_{4}(\widetilde{K},K)]\}.
\end{align}
and $\Omega\Gamma^{0}(\widetilde{K},K)+i\mathrm{div}_{\mb{q}}\pmb\Gamma=$
\begin{align}&\Omega+i\mathrm{div}_{\mb{q}}\pmb\gamma
+\Delta^{2}_{1}G_{1,1}(-K)G_{1,1}(-\widetilde{K})\nonumber\\&\times\{[\Omega+i\mathrm{div}_{\mb{q}}\pmb\gamma_{1}(-K,-\widetilde{K})]
\nonumber\\&+G_{0,4}(K)G_{0,4}(\widetilde{K})[\Omega+i\mathrm{div}_{\mb{q}}\pmb\gamma_{4}(\widetilde{K},K)] \}.\end{align}

In this form we can then solve for the exact full vertex 
\begin{align}
&\Gamma^{\mu}(\widetilde{K},K)=\gamma^{\mu}(\widetilde{K},K)+\Delta^{2}_{1}G_{1,1}(-K)G_{1,1}(-\widetilde{K})\nonumber\\&\times
[\gamma^{\mu}_{1}(-K,-\widetilde{K})+\Delta^{2}_{2}G_{0,4}(K)G_{0,4}(\widetilde{K})\gamma^{\mu}_{4}(\widetilde{K},K)]\nonumber\\
&+\Delta^{2}_{2}G_{1,2}(-K)G_{1,2}(-\widetilde{K})\nonumber\\&\times[\gamma^{\mu}_{2}(-K,-\widetilde{K})
+\Delta^{2}_{1}G_{0,3}(K)G_{0,3}(\widetilde{K})\gamma^{\mu}_{3}(\widetilde{K},K)].
\label{eq:V}\end{align}
Here we have now included the second term from $\Sigma_{pg}(K)$ in
Eq.~(\ref{eq:LSE}).

We emphasize this is not an expansion in the bare vertices. Rather, the WTI is used to obtain the exact full vertex.
The crucial step is that the self energy does not depend
on the full Green's function. If it did, then the full vertex would appear on both sides of the equation, 
reducing the problem to a Bethe-Salpeter equation \cite{Schrieffersbook}.

Using the full vertex, the exact response function can then be determined via Eq. (\ref{eq:RF}). 
The Amperean pairing response functions have twenty one associated 
Feynman diagrams if one considered the charge density wave: one of one loop order
(two Green's functions), four of two loop order (four Green's
functions), and four of three loop order (six Green's functions),
plus an additional twelve diagrams with an odd number of Green's functions.
The twenty one Feynman diagrams contributing to the response functions are 
presented in the Supplemental Material.

The bare vertices for the density component are given by
$\gamma^{0}(\widetilde{K},K)=\gamma^{0}_{i}(\widetilde{K},K)=1$. 
This then allows the exact density-density
response function $P_{\rho\rho}(Q)$ to be computed for all $Q$.
More complicated, for an arbitrary band-structure, are the bare vertices
that enter into the current-current correlation function.
However, in the limit $q\rightarrow0$
these can be determined from Eq.~(\ref{eq:RF02}).
The same reasoning which is used to determine 
$P_{\rho\rho}(Q)$ for all $Q$ is applicable to
the spin (density) response, as measured in neutron experiments.

\begin{figure*}
\subfloat{\includegraphics[height=3.9cm,clip]{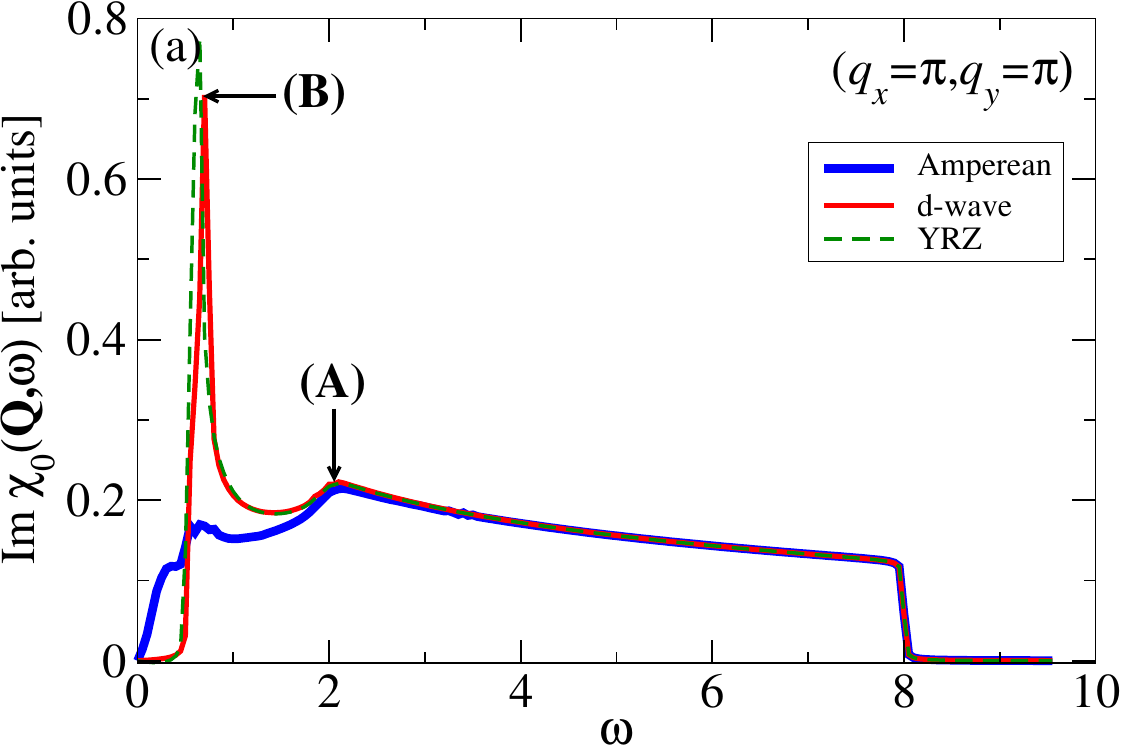}}\hspace{1mm}
\subfloat{\includegraphics[height=3.9cm,clip]{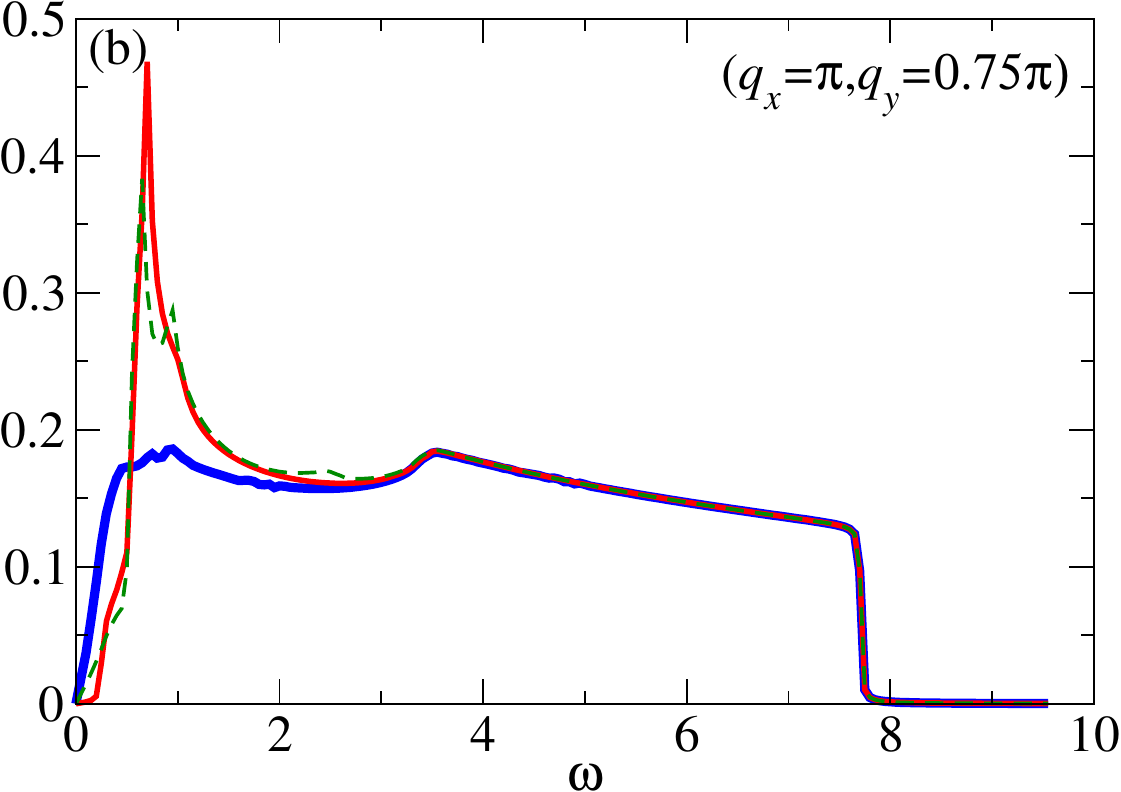}}\hspace{1mm}
\subfloat{\includegraphics[height=3.9cm,clip]{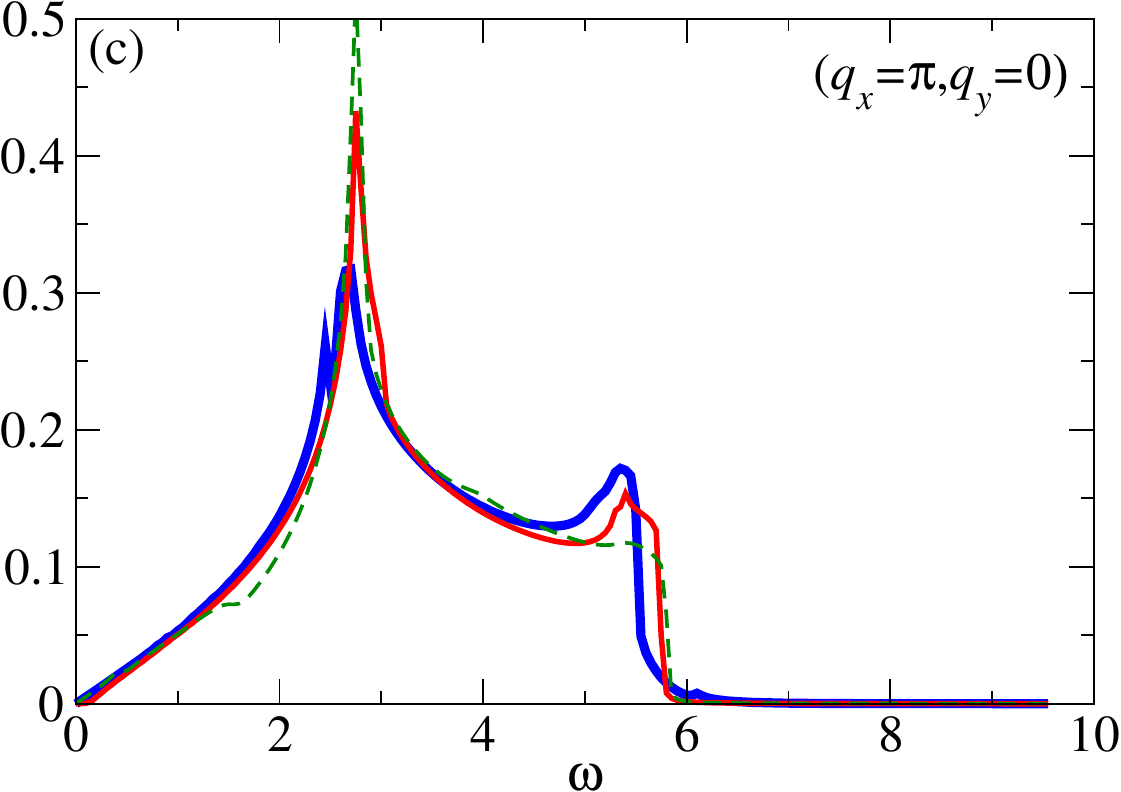}}
\caption{Comparison of the spin density
correlation function $\mathrm{Im}\chi_{0}(Q)=-\mathrm{Im}P^{00}_{S}(\mb{q},\omega)$ 
for three different values of $\mb{q}$ in the Amperean, $d$-wave and YRZ pseudogap theories. 
In (a) we have labeled the Van Hove peaks appearing in the $d$-wave theory, 
which appear as saddle points in the contour plot of Fig. (\ref{fig:contour}).
Here we use the band structure given in \cite{Comin} with $T=0.01$ and 
a broadening of $\Gamma=0.01$. The doping $p=0.12$ and the 
chemical potential $\mu$ is fixed by the Luttinger sum rule. 
The band-structure and frequency are all normalized by $t$, and
the gap function has an amplitude of $\Delta_{0}=0.15$. 
For the Amperean theory we use a $k_{x}, k_{y}$-symmetrized
Gaussian \cite{Lee_2014} gap function.}
\label{fig:Pplot}
\end{figure*}

\begin{figure}
\subfloat{\includegraphics[scale=0.88,clip]{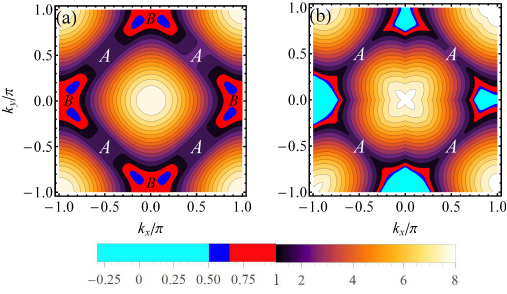}}
\caption{The equal energy contours 
$E_2(\mb{k}) \equiv E_{\mb{k}} + E_{\mb{k+q}}$
which appear as the integrand in $\mathrm{Im}\chi_{0}(\mb{q},\omega)$
for both (a) $\it d$-wave and 
(b) the Amperean pseudogap schemes. Here $\mb{q}=(\pi,\pi)$. Note there are
several energy scales as indicated by the legend. The labels $A$ 
and $B$ indicate the location of the 
saddle points of $\mathrm{Im}\chi_{0}(\mb{q},\omega)$.}
\label{fig:contour}
\end{figure}

The full spin response function $P^{\mu\nu}_{S}(Q)$ is defined by
\begin{align}P^{\mu\nu}_{S}(Q)=\sum_{K}\sum_{\sigma}\Gamma^{\mu}_{S_{\sigma}}(\widetilde{K},K)G(K)
\gamma^{\nu}_{S_{\sigma}}(K,\widetilde{K})G(\widetilde{K}).\end{align}   
Here the bare spin vertex is denoted by $\gamma^{\mu}_{S_{\sigma}}(\widetilde{K},K)$, 
where $S_{\sigma}=\pm1$ and $S_{\bar{\sigma}}=-S_{\sigma}$.
The bare WTI for the spin vertex is
$\Omega+i\mathrm{div}_{\mb{q}}\pmb\gamma_{S_{\sigma}}=S_{\sigma}(G^{-1}_{0}(\widetilde{K})-G^{-1}_{0}(K)) 
=S_{\sigma}(\Omega-\xi_{\mb{k+q}}+\xi_{\mb{k}})$.
Similarly the full WTI for the full spin vertex 
$\Gamma^{\mu}_{S_{\sigma}} 
(\widetilde{K},K)$ is                                                                           
\begin{equation}\Omega\Gamma^{0}_{S_{\sigma}}+i\mathrm{div}_{\mb{q}}\pmb\Gamma_{S_{\sigma}}=S
_{\sigma}(G^{-1}(\widetilde{K})-G^{-1}(K)).\end{equation} 
We can then read off the spin-spin correlation function directly using Eq. (\ref{eq:V}).

From the established constraints
on the bare and full vertices one can directly derive \cite{Tremblay} the 
$f$-sum rule for the density-density and spin density response functions:
\begin{equation}\label{eq:PRR}\int\frac{d\omega}{\pi}(-\omega\mathrm{Im}P^{00}(Q))
=2\sum_{\mb{k}}n_{\mb{k}}[\xi_{\mb{k+q}}+\xi_{\mb{k-q}}-2\xi_{\mb{k}}],\end{equation}
where $n_{\mb{k}}=T\sum_{i\omega}G(K).$ Importantly, this sum rule (and counterparts for the 
current-current correlation function) are
satisfied exactly providing the response functions 
are consistent with the WTI.
This is discussed in more detail in the Supplemental Material.

\textit{Behavior of the neutron cross section: 
Comparison of Pseudogap theories.$-$}
For illustrative purposes
we focus on the spin-density response function, conventionally called $\chi_0(Q)$.
Importantly, ensuring Eq. (\ref{eq:PRR}) is satisfied 
provides tight control over numerical calculations of this correlation function.
When no simplifications are introduced,
our numerical calculations agree with the sum rule to an accuracy of the order of $0.1-0.2$ percent \cite{discuss} in all three models.
The quantity $\mathrm{Im}\chi_0(Q)\equiv-\mathrm{Im}P^{00}(Q)$
is the theoretical basis for neutron scattering experiments. 
[Note we adopt the sign convention for the density correlation functions,
$P_{\rho\rho}(Q)=P^{00}(Q)$ for spin and charge.]

For simple $d$-wave pairing models,
a very reasonable comparison between theory and neutron data has been reported
at high temperatures (where one sees a reflection of the
fermiology \cite{Qimiao1,Zha1}) and below $T_{C}$ (where one sees both commensurate 
($\pi,\pi$) \cite{Liu2} and slightly incommensurate 
frequency dependent ``hourglass" structure \cite{Zha2,Kao}). 
This approach to neutron
scattering presents a (rather successful) rival scheme to stripe approaches;
many different theories, built on different microscopics,
have arrived at similar behavior 
\cite{Lavagna,LeeBrinckmann2,Normanneutron,RiceMag1}.
In the pseudogap phase (which has received less attention
theoretically), there are peaks at and near ($\pi,\pi$)
\cite{Birg,Dai,Keimer} which have been recently argued \cite{Keimer}
to reflect some degree of broken orientational symmetry.

Here we compare the results for $\chi_0(Q)$ using
three different theories of the pseudogap: 
a simple $d$-wave pseudogap, the theory of
Yang, Rice and Zhang, and that of Amperean pairing.
For the Amperean case we follow \cite{Lee_2014} and consider
the simpler $3 \times 3$ reduced Hamiltonian.
In this $3 \times 3$ form, $C_1 =C_2 =0$ and terms involving
$\xi_{\mb{k} \pm2\mb{p}}$ are dropped.
We do not include the effects of the widely used RPA enhanced
denominator introduced in \cite{Si2}.
In the RPA enhanced form $\chi(Q) = \chi_{0}(Q)/[1+ J(Q) \chi_{0}(Q)]$, 
where $J(Q)$ is an effective exchange. 
Even though $\chi_{0}(Q)$ is exact, introducing
this ratio will lead to a violation of the $f$-sum rule; this effect
is not central to distinguishing between theories, as is our goal here.

Figure (\ref{fig:Pplot}) presents a plot of $\mathrm{Im}\chi_{0}(Q)$, 
for three fixed $\mb{q}$ corresponding to ($\pi, \pi)$ in (a), ($\pi, 0.75\pi$) in (b) and 
($\pi, 0$) in (c) as functions of $\omega$. 
The normal state (above $T^*$) band-structure is taken to be
the same, as is the pseudogap amplitude. 
The behavior in the low $\omega$ regime
is principally, but not exclusively, dominated by the effects of
the gap, while at very high $\omega$ the behavior is
band-structure dominated. Importantly, all theories essentially
converge once $\omega$ is much larger than the gap.
This means that interesting
effects associated with high energy
scales \cite{Keimerparamag} such as observed in recent
RIXS experiments, would not be specific to a given theory.

Figure (\ref{fig:Pplot}) shows that there is
little difference in the spin dynamics 
between the approach of YRZ \cite{YRZ}
and that of a $d$-wave pseudogap, emphasized
earlier in a different context \cite{Peter1}
and helps to explain the
literature claims of successful reconciliation
with the data that surround both scenarios \cite{RiceMag1,Zha2,Kao}.

What appears most distinctive is the Amperean pairing
response function, particularly away 
from $\mb{q} = (\pi, 0)$.
Notable here is the absence of
the sharp Van Hove peak (marked by $B$ in Fig. (\ref{fig:Pplot}))
which appears in both other theories
and which is ultimately responsible for commensurate
peaks or neutron resonance effects \cite{Liu2}.
Also missing from the Amperean scenario
is the so-called spin-gap, apparent at $(\pi, \pi)$
in both the other two theories. 
Rather, for Amperean pairing there are
multiple low energy processes which contribute to the 
spin density correlation function.

To better understand these processes, 
in Fig.~(\ref{fig:contour}) we probe the dominant component
of the integrand in $\mathrm{Im}\chi_{0}(Q)$ 
near $\mb{q} = (\pi, \pi)$ for the Amperean (right) as compared with
$d$-wave pseudogap (left) scenarios.
We show the equal energy contours 
for the sum of the quasi-particle
dispersions: $E_2(\mb{k}) \equiv E_{\mb{k}} + E_{\mb{k+q}}$,
vs $k_x$ and $k_y$ in the pseudogap state \cite{e2k}.
Indicated on the figure are the Van Hove singularities $A$ and $B$ (saddle points in 
the contour plot), as labeled in Fig. (\ref{fig:Pplot}a). 
The lower energy Van Hove point (point $B$)
is clearly suppressed in the Amperean pairing case,
while it is very pronounced and found
to be important \cite{Kao} for the $d$-wave case.
Also evident from the cyan region in Fig.~(\ref{fig:contour}) 
is the absence of a low $\omega$ minimum (spin gap) in
$E_2({\bf k})$, as found in both the other two theories, as
well as in the integrated response function.

\textit{Conclusions.$-$}
The central contribution of this paper has been to establish
an analytically and numerically controlled methodology
for addressing the long list of two particle cuprate measurements.
Given a mean field like self energy,
the exploitation of the Ward Takahashi identity (and related sum rules)
allows one to evaluate two particle properties, and in this
way achieve the same level of accuracy in these comparisons, as in, say, ARPES.
To demonstrate the utility of this method, 
we address the spin density response functions of neutron scattering and
have singled out signatures of the recently proposed
Amperean pairing theory \cite{Lee_2014}.
We cannot firmly establish that this pair density
wave theory is inconsistent with experiments (without digressing
from our goals and including
the sum-rule-inconsistent RPA enhancement denominator \cite{Si2}), but it does 
lead to a rather featureless neutron cross section \cite{generality}.
We report two distinctive observations:
the absence of both spin gap effects and of
the sharp Van Hove peaks near $(\pi,\pi)$.

This work is supported by NSF-MRSEC Grant 0820054.
We are grateful to A.-M. S. Tremblay, Yan He and Adam Ran\c{c}on for helpful conversations. 
\bibliographystyle{apsrev}
\bibliography{Review}

\begin{thebibliography}{35}
\bibitem[{dis()}]{discuss}
\bibinfo{note}{For simplicity in the text, we will ignore terms that arise from
  the wave vector dependence of the $d$-wave gap function. In general this is a
  small effect. We can see via the sum rule accuracy the importance of
  including the full wave vector dependence of the $d$-wave gap. When the wave
  vector dependence of the gap is ignored, the sum rule accuracy is still very
  good, but now of the order of 2-3 percent.}
\end{thebibliography}

\clearpage


\section*{Supplementary material}

# Supplemental Material: Exact correlation functions in the cuprate pseudogap phase: combined effects of charge order and pairing

Rufus Boyack, Chien-Te Wu, Peter Scherpelz, and K. Levin
*James Franck Institute, University of Chicago, Chicago, Illinois 60637, USA*

## I. INCLUSION OF BOTH CHARGE-DENSITY WAVE AND PAIR-DENSITY WAVE EFFECTS

Now we extend our results to include the effects of the charge-density waves $C_1$ and $C_2$: $C_1, C_2 \neq 0$. The full self energy, as in Eq. (1) of the main text, is given by

$$\Sigma_{pg}(K) = \frac{\Delta_1^2}{\omega + \xi_{\mathbf{k-p}} - \frac{\Delta_2^2}{\omega - \xi_{\mathbf{k-2p}}}} + \frac{\Delta_2^2}{\omega + \xi_{\mathbf{k+p}} - \frac{\Delta_1^2}{\omega - \xi_{\mathbf{k+2p}}}} + \frac{C_1^2}{\omega - \xi_{\mathbf{k+2p}} - \frac{\Delta_1^2}{\omega + \xi_{\mathbf{k+p}}}} + \frac{C_2^2}{\omega - \xi_{\mathbf{k-2p}} - \frac{\Delta_2^2}{\omega + \xi_{\mathbf{k-p}}}} + \frac{2\Delta_1 \Delta_2 C_1}{(\omega - \xi_{\mathbf{k+2p}})(\omega + \xi_{\mathbf{k+p}}) - \Delta_1^2} + \frac{2\Delta_1 \Delta_2 C_2}{(\omega - \xi_{\mathbf{k-2p}})(\omega + \xi_{\mathbf{k-p}}) - \Delta_2^2}. \quad (1)$$

For convenience we define the following four bare (inverse) Green's functions $G_{0,i}^{-1}(K) = (\omega - \xi_{\mathbf{k},i}), i \in \{1,2,3,4\}$, where $\xi_{\mathbf{k},1} = \xi_{\mathbf{k+p}}, \xi_{\mathbf{k},2} = \xi_{\mathbf{k-p}}, \xi_{\mathbf{k},3} = \xi_{\mathbf{k+2p}}, \xi_{\mathbf{k},4} = \xi_{\mathbf{k-2p}}$ are four dispersion relations. (The usual bare inverse Green's function is denoted by $G_0^{-1}(K) = \omega - \xi_{\mathbf{k}} = \omega - \epsilon_{\mathbf{k}} + \mu$.) Using these definitions, we can then define the following partially dressed Green's functions:

$$G_{1,1}^{-1}(K) = \omega - \xi_{\mathbf{k},1} - \frac{\Delta_2^2}{\omega + \xi_{\mathbf{k},3}} = G_{0,1}^{-1}(K) + \Delta_2^2 G_{0,4}(-K), \quad (2)$$

$$G_{1,2}^{-1}(K) = \omega - \xi_{\mathbf{k},2} - \frac{\Delta_1^2}{\omega + \xi_{\mathbf{k},4}} = G_{0,2}^{-1}(K) + \Delta_1^2 G_{0,3}(-K), \quad (3)$$

$$G_{1,3}^{-1}(K) = \omega - \xi_{\mathbf{k},3} - \frac{\Delta_1^2}{\omega + \xi_{\mathbf{k},1}} = G_{0,3}^{-1}(K) + \Delta_1^2 G_{0,2}(-K), \quad (4)$$

$$G_{1,4}^{-1}(K) = \omega - \xi_{\mathbf{k},4} - \frac{\Delta_2^2}{\omega + \xi_{\mathbf{k},2}} = G_{0,4}^{-1}(K) + \Delta_2^2 G_{0,1}(-K). \quad (5)$$

The full self energy of the Amperean pairing theory is then

$$\begin{aligned}\Sigma_{pg}(K) = & -\Delta_1^2 G_{1,1}(-K) - \Delta_2^2 G_{1,2}(-K) + C_1^2 G_{1,3}(K) + C_2^2 G_{1,4}(K) \\ & - \Delta_1 \Delta_2 C_1 G_{0,2}(-K) G_{1,3}(K) - \Delta_1 \Delta_2 C_1 G_{0,3}(K) G_{1,2}(-K) \\ & - \Delta_1 \Delta_2 C_2 G_{0,1}(-K) G_{1,4}(K) - \Delta_1 \Delta_2 C_2 G_{0,4}(K) G_{1,1}(-K). \end{aligned} \quad (6)$$

Note that $G_{0,4}(K) G_{1,1}(-K) = G_{0,1}(-K) G_{1,4}(K)$ and $G_{0,3}(K) G_{1,2}(-K) = G_{0,2}(-K) G_{1,3}(K)$; thus the two terms on the second and third lines can each be combined into a single expression. However, from here on out we will use the symmetric form of the self energy as given in Eq. (6).

In order to derive the full vertex $\Gamma^\mu(\widetilde{K}, K)$, where $\widetilde{K} \equiv K + Q$, associated with the self energy in Eq. (1) and Eq. (6), we use the Ward-Takahashi identity (WTI). The WTI, for the vertex $\Gamma^\mu = (\Gamma^0, \boldsymbol{\Gamma})$, on a lattice is

$$\begin{aligned}\Omega \Gamma^0(\widetilde{K}, K) + i \text{div}_{\mathbf{q}} \boldsymbol{\Gamma}(\widetilde{K}, K) &= G^{-1}(\widetilde{K}) - G^{-1}(K), \\ &= \Omega + i \text{div}_{\mathbf{q}} \boldsymbol{\gamma}(\widetilde{K}, K) + \Sigma_{pg}(K) - \Sigma_{pg}(\widetilde{K}). \end{aligned} \quad (7)$$

Here we have used Dyson's equation $G^{-1}(K) = G_0^{-1}(K) - \Sigma_{pg}(K)$, along with the bare WTI: $\Omega \gamma^0(\widetilde{K}, K) + i \text{div}_{\mathbf{q}} \boldsymbol{\gamma}(\widetilde{K}, K) = G_0^{-1}(\widetilde{K}) - G_0^{-1}(K)$. Since the self energy contains only bare and partially dressed Green's functions, the WTI allows the full vertex $\Gamma^\mu(\widetilde{K}, K)$ to be obtained explicitly. To show this, we need to compute the difference $\Sigma_{pg}(K) - \Sigma_{pg}(\widetilde{K})$ appearing in Eq. (7).

Using the self energy in Eq. (6), we find that

$$\Sigma_{pg}(K) - \Sigma_{pg}(\widetilde{K}) =$$
$$\Delta_1^2 G_{1,1}(-K) \left[ G_{1,1}^{-1}(-K) - G_{1,1}^{-1}(-\widetilde{K}) \right] G_{1,1}(-\widetilde{K})$$
$$+\Delta_2^2 G_{1,2}(-K) \left[ G_{1,2}^{-1}(-K) - G_{1,2}^{-1}(-\widetilde{K}) \right] G_{1,2}(-\widetilde{K})$$
$$+C_1^2 G_{1,3}(\widetilde{K}) \left[ G_{1,3}^{-1}(\widetilde{K}) - G_{1,3}^{-1}(K) \right] G_{1,3}(K)$$
$$+C_2^2 G_{1,4}(\widetilde{K}) \left[ G_{1,4}^{-1}(\widetilde{K}) - G_{1,4}^{-1}(K) \right] G_{1,4}(K)$$
$$+\Delta_1 \Delta_2 C_1 G_{1,3}(K) G_{0,2}(-K) \left[ G_{0,2}^{-1}(-K) - G_{0,2}^{-1}(-\widetilde{K}) \right] G_{0,2}(-\widetilde{K})$$
$$-\Delta_1 \Delta_2 C_1 G_{0,2}(-\widetilde{K}) G_{1,3}(\widetilde{K}) \left[ G_{1,3}^{-1}(\widetilde{K}) - G_{1,3}^{-1}(K) \right] G_{1,3}(K)$$
$$+\Delta_1 \Delta_2 C_1 G_{1,2}(-K) \left[ G_{1,2}^{-1}(-K) - G_{1,2}^{-1}(-\widetilde{K}) \right] G_{1,2}(-\widetilde{K}) G_{0,3}(\widetilde{K})$$
$$-\Delta_1 \Delta_2 C_1 G_{0,3}(\widetilde{K}) \left[ G_{0,3}^{-1}(\widetilde{K}) - G_{0,3}^{-1}(K) \right] G_{0,3}(K) G_{1,2}(-K)$$
$$+\Delta_1 \Delta_2 C_2 G_{1,4}(K) G_{0,1}(-K) \left[ G_{0,1}^{-1}(-K) - G_{0,1}^{-1}(-\widetilde{K}) \right] G_{0,1}(-\widetilde{K})$$
$$-\Delta_1 \Delta_2 C_2 G_{0,1}(-\widetilde{K}) G_{1,4}(\widetilde{K}) \left[ G_{1,4}^{-1}(\widetilde{K}) - G_{1,4}^{-1}(K) \right] G_{1,4}(K)$$
$$+\Delta_1 \Delta_2 C_2 G_{1,1}(-K) \left[ G_{1,1}^{-1}(-K) - G_{1,1}^{-1}(-\widetilde{K}) \right] G_{1,1}(-\widetilde{K}) G_{0,4}(\widetilde{K})$$
$$-\Delta_1 \Delta_2 C_2 G_{0,4}(\widetilde{K}) \left[ G_{0,4}^{-1}(\widetilde{K}) - G_{0,4}^{-1}(K) \right] G_{0,4}(K) G_{1,1}(-K). \tag{8}$$

The terms in square brackets can all be expressed as contractions of various bare vertices, by using the bare Ward-Takahashi identities. For example, using Eqs. (2-5), we can write:

$$G_{1,1}^{-1}(-K) - G_{1,1}^{-1}(-\widetilde{K}) = G_{0,1}^{-1}(-K) - G_{0,1}^{-1}(-\widetilde{K}) + \Delta_2^2 G_{0,4}(K) G_{0,4}(\widetilde{K}) \left[ G_{0,4}^{-1}(\widetilde{K}) - G_{0,4}^{-1}(K) \right]$$
$$= \Omega \gamma_1^0(-K, -\widetilde{K}) + i \mathrm{div}_{\mathbf{q}} \boldsymbol{\gamma}_1(-K, -\widetilde{K}) + \Delta_2^2 G_{0,4}(\widetilde{K}) \left[ \Omega \gamma_4^0(\widetilde{K}, K) + i \mathrm{div}_{\mathbf{q}} \boldsymbol{\gamma}_4(\widetilde{K}, K) \right] G_{0,4}(K). \tag{9}$$

$$G_{1,2}^{-1}(-K) - G_{1,2}^{-1}(-\widetilde{K}) = G_{0,2}^{-1}(-K) - G_{0,2}^{-1}(-\widetilde{K}) + \Delta_1^2 G_{0,3}(K) G_{0,3}(\widetilde{K}) \left[ G_{0,3}^{-1}(\widetilde{K}) - G_{0,3}^{-1}(K) \right]$$
$$= \Omega \gamma_2^0(-K, -\widetilde{K}) + i \mathrm{div}_{\mathbf{q}} \boldsymbol{\gamma}_2(-K, -\widetilde{K}) + \Delta_1^2 G_{0,3}(\widetilde{K}) \left[ \Omega \gamma_3^0(\widetilde{K}, K) + i \mathrm{div}_{\mathbf{q}} \boldsymbol{\gamma}_3(\widetilde{K}, K) \right] G_{0,3}(K). \tag{10}$$

$$G_{1,3}^{-1}(\widetilde{K}) - G_{1,3}^{-1}(K) = G_{0,3}^{-1}(\widetilde{K}) - G_{0,3}^{-1}(K) + \Delta_1^2 G_{0,2}(-K) G_{0,2}(-\widetilde{K}) \left[ G_{0,2}^{-1}(-K) - G_{0,2}^{-1}(-\widetilde{K}) \right]$$
$$= \Omega \gamma_3^0(\widetilde{K}, K) + i \mathrm{div}_{\mathbf{q}} \boldsymbol{\gamma}_3(\widetilde{K}, K) + \Delta_1^2 G_{0,2}(-K) \left[ \Omega \gamma_2^0(-K, -\widetilde{K}) + i \mathrm{div}_{\mathbf{q}} \boldsymbol{\gamma}_2(-K, -\widetilde{K}) \right] G_{0,2}(-\widetilde{K}). \tag{11}$$

$$G_{1,4}^{-1}(\widetilde{K}) - G_{1,4}^{-1}(K) = G_{0,4}^{-1}(\widetilde{K}) - G_{0,4}^{-1}(K) + \Delta_2^2 G_{0,1}(-K) G_{0,1}(-\widetilde{K}) \left[ G_{0,1}^{-1}(-K) - G_{0,1}^{-1}(-\widetilde{K}) \right]$$
$$= \Omega \gamma_4^0(\widetilde{K}, K) + i \mathrm{div}_{\mathbf{q}} \boldsymbol{\gamma}_4(\widetilde{K}, K) + \Delta_2^2 G_{0,1}(-K) \left[ \Omega \gamma_1^0(-K, -\widetilde{K}) + i \mathrm{div}_{\mathbf{q}} \boldsymbol{\gamma}_1(-K, -\widetilde{K}) \right] G_{0,1}(-\widetilde{K}). \tag{12}$$

In the last step in each of these expressions we have used the bare WTI's. This procedure reduces the terms in square brackets in Eq. (8) to a contraction of the time component and vector component of a bare vertex. Since the full WTI involves a similar contraction of the time component and vector component of the full vertex, we can assert then that the individual vertex that appears above is a contribution to the full vertex. Performing this procedure for all terms in Eq. (8) then allows us to extract the full vertex via Eq. (7).

This gives the full vertex as

$$\begin{aligned}\Gamma^\mu(\widetilde{K},K) =& \gamma^\mu(\widetilde{K},K) \\
&+ \Delta_1^2 G_{1,1}(-K)\left[\gamma_1^\mu(-K,-\widetilde{K}) + \Delta_2^2 G_{0,4}(\widetilde{K})\gamma_4^\mu(\widetilde{K},K)G_{0,4}(K)\right]G_{1,1}(-\widetilde{K}) \\
&+ \Delta_2^2 G_{1,2}(-K)\left[\gamma_2^\mu(-K,-\widetilde{K}) + \Delta_1^2 G_{0,3}(\widetilde{K})\gamma_3^\mu(\widetilde{K},K)G_{0,3}(K)\right]G_{1,2}(-\widetilde{K}) \\
&+ C_1^2 G_{1,3}(\widetilde{K})\left[\gamma_3^\mu(\widetilde{K},K) + \Delta_1^2 G_{0,2}(-K)\gamma_2^\mu(-K,-\widetilde{K})G_{0,2}(-\widetilde{K})\right]G_{1,3}(K) \\
&+ C_2^2 G_{1,4}(\widetilde{K})\left[\gamma_4^\mu(\widetilde{K},K) + \Delta_2^2 G_{0,1}(-K)\gamma_1^\mu(-K,-\widetilde{K})G_{0,1}(-\widetilde{K})\right]G_{1,4}(K) \\
&+\Delta_1\Delta_2 C_1 G_{1,3}(K)G_{0,2}(-K)\gamma_2^\mu(-K,-\widetilde{K})G_{0,2}(-\widetilde{K}) \\
&-\Delta_1\Delta_2 C_1 G_{0,2}(-\widetilde{K})G_{1,3}(\widetilde{K})\left[\gamma_3^\mu(\widetilde{K},K) + \Delta_1^2 G_{0,2}(-K)\gamma_2^\mu(-K,-\widetilde{K})G_{0,2}(-\widetilde{K})\right]G_{1,3}(K) \\
&+\Delta_1\Delta_2 C_1 G_{1,2}(-K)\left[\gamma_2^\mu(-K,-\widetilde{K}) + \Delta_1^2 G_{0,3}(\widetilde{K})\gamma_3^\mu(\widetilde{K},K)G_{0,3}(K)\right]G_{1,2}(-\widetilde{K})G_{0,3}(\widetilde{K}) \\
&-\Delta_1\Delta_2 C_1 G_{0,3}(\widetilde{K})\gamma_3^\mu(\widetilde{K},K)G_{0,3}(K)G_{1,2}(-K) \\
&+\Delta_1\Delta_2 C_2 G_{1,4}(K)G_{0,1}(-K)\gamma_1^\mu(-K,-\widetilde{K})G_{0,1}(-\widetilde{K}) \\
&-\Delta_1\Delta_2 C_2 G_{0,1}(-\widetilde{K})G_{1,4}(\widetilde{K})\left[\gamma_4^\mu(\widetilde{K},K) + \Delta_2^2 G_{0,1}(-K)\gamma_1^\mu(-K,-\widetilde{K})G_{0,1}(-\widetilde{K})\right]G_{1,4}(K) \\
&+\Delta_1\Delta_2 C_2 G_{1,1}(-K)\left[\gamma_1^\mu(-K,-\widetilde{K}) + \Delta_2^2 G_{0,4}(\widetilde{K})\gamma_4^\mu(\widetilde{K},K)G_{0,4}(K)\right]G_{1,1}(-\widetilde{K})G_{0,4}(\widetilde{K}) \\
&-\Delta_1\Delta_2 C_2 G_{0,4}(\widetilde{K})\gamma_4^\mu(\widetilde{K},K)G_{0,4}(K)G_{1,1}(-K). \end{aligned} \quad (13)$$

Given the full and bare vertices $\Gamma^\mu, \gamma^\nu$, the full response functions are then

$$P^{\mu\nu}(Q) = 2\sum_K G(\widetilde{K})\Gamma^\mu(\widetilde{K},K)G(K)\gamma^\nu(K,\widetilde{K}). \quad (14)$$

There are twenty one Feynman diagrams that contribute to the response functions, which are shown in Figure (1). In the more general case of wave vector dependent gaps, there are additional terms involving the derivatives of the gap, which, in the 3×3 reduced Hamiltonian theory give an essentially negligible contribution[1]. The response functions are completely specified once the various bare vertices are given. We have to rely on the semi-classical approximation to obtain the form of the bare vertices associated with the current. In the limit $\mathbf{q} \to 0$, this approximation becomes rigorously correct.

## II. RELATION BETWEEN SUM RULES AND THE WARD-TAKAHASHI IDENTITY

### A. $f$ sum rule

The exact response functions are determined from Eq. (14). (We take the density-density correlation function to be $P_{\rho\rho}(Q) = P^{00}(Q)$ and the current-current correlation function is $\overleftrightarrow{P}(Q) = P^{ij}(Q)$, $i,j \in \{1,2,3\}$.) In the main text we assert that, given bare and full vertices that satisfy the associated WTI, the following $f$-sum rule is satisfied

$$\int \frac{d\omega}{\pi}(-\omega \text{Im} P^{00}(Q)) = 2\sum_\mathbf{k} n_\mathbf{k}(\xi_{\mathbf{k}+\mathbf{q}} + \xi_{\mathbf{k}-\mathbf{q}} - 2\xi_\mathbf{k}), \quad (15)$$

where $n_\mathbf{k} = T\sum_{i\omega} G(K)$. This sum rule can be established as follows.

The WTI for the full and bare vertices[2], $\Gamma^\mu(\widetilde{K},K) = (\Gamma^0(\widetilde{K},K), \mathbf{\Gamma}(\widetilde{K},K))$ and $\gamma^\mu(\widetilde{K},K) = (1, \boldsymbol{\gamma}(\widetilde{K},K))$ are

$$\begin{aligned}\Omega\Gamma^0 + i\text{div}_\mathbf{q}\mathbf{\Gamma} &= G^{-1}(\widetilde{K}) - G^{-1}(K), \\
\Omega + i\text{div}_\mathbf{q}\boldsymbol{\gamma} &= G_0^{-1}(\widetilde{K}) - G_0^{-1}(K) = \Omega - \xi_{\mathbf{k}+\mathbf{q}} + \xi_\mathbf{k}.\end{aligned} \quad (16)$$

The response kernel is defined as $K^{\mu\nu}(Q) = P^{\mu\nu}(Q) + \left(\frac{n}{m}\right)^{\mu\nu}_{\text{dia}}(1 - \delta_{0,\nu}\delta_{\mu,\nu})$, where there is no summation over indices in the second term. Applying the WTI to $P^{\mu 0}$ while setting $\nu = 0$ the result is

$$\Omega P^{00} + i\text{div}_\mathbf{q} P^{i0} = 2\sum_K G(\widetilde{K})G(K)[G^{-1}(\widetilde{K}) - G^{-1}(K)] = 0. \quad (17)$$

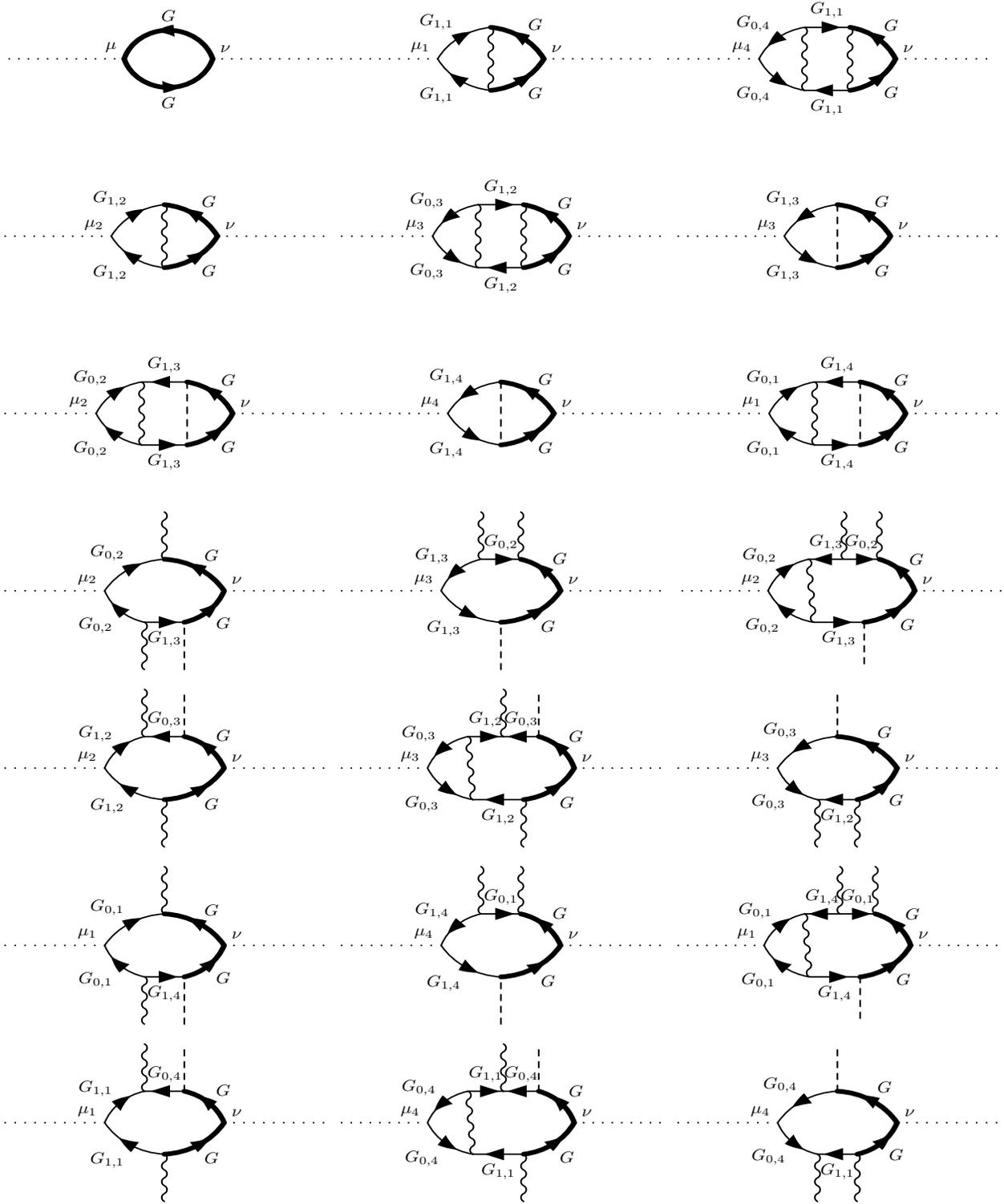

FIG. 1. All twenty one Feynman diagrams which contribute to the response functions $P^{\mu\nu}(Q)$. The wavy lines denote either $\Delta_1$ or $\Delta_2$, whereas the dashed lines denote either $C_1$ or $C_2$. The various Green's function are labeled. The order of the Feynman diagrams, from left to right and top to bottom, corresponds to the terms appearing in Eq. (13).

If we set $\nu = j = \{1, 2, 3\}$ and apply the WTI to $P^{\mu j}$ the result is

$$\Omega P^{0j} + i\text{div}_{\mathbf{q}} P^{ij} = 2 \sum_{K} G(K)[\boldsymbol{\gamma}(K, \widetilde{K}) - \boldsymbol{\gamma}(K - Q, K)]. \tag{18}$$

Setting $\Omega = 0$ and then operating with $i\text{div}_{\mathbf{q}}$ gives

$$i\text{div}_{\mathbf{q}} i\text{div}_{\mathbf{q}} P^{ij}(\mathbf{q}, 0) = 2 \sum_{\mathbf{k}} n_{\mathbf{k}}[2\xi_{\mathbf{k}} - \xi_{\mathbf{k}+\mathbf{q}} - \xi_{\mathbf{k}-\mathbf{q}}]. \tag{19}$$

Now use the identity $\text{Im} P^{i0}(\mathbf{q}, \omega) = -\text{Im} P^{0i}(-\mathbf{q}, -\omega)$ and Eq. (17), Eq. (18) and Eq. (19) to solve for $\text{Im} P^{00}$ in terms of $\text{Im} P^{ij}$. By applying the Kramers-Kronig relations we then have the sum rule in Eq. (15). Importantly, we have proved the $f$-sum rule for all values of $\mathbf{q}$. This proof depends on having bare and full vertices $\gamma^{\mu}(\widetilde{K}, K)$ and $\Gamma^{\mu}(\widetilde{K}, K)$ which satisfy the Ward Takahashi identity.

### B. Longitudinal sum rule

By using the relationship between $\text{Im} P^{ij}$ and $\text{Im} P^{00}$, along with the $f$-sum rule, the longitudinal sum rule is obtained. The abstract form of the longitudinal sum rule is

$$\int \frac{d\omega}{\pi} \left( -\frac{\text{Im}\{i\text{div}_{\mathbf{q}} i\text{div}_{\mathbf{q}} \overset{\leftrightarrow}{P}(Q)\}}{\omega} \right) = 2 \sum_{\mathbf{k}} n_{\mathbf{k}}(\xi_{\mathbf{k}+\mathbf{q}} + \xi_{\mathbf{k}-\mathbf{q}} - 2\xi_{\mathbf{k}}). \tag{20}$$

The WTI for the bare vertices implies that $i\text{div}_{\mathbf{q}} \gamma = \xi_{\mathbf{k}} - \xi_{\mathbf{k}+\mathbf{q}}$. Thus by applying this identity to the vertices that appear in Eq. (14), the longitudinal sum rule is manifestly satisfied. While the response functions can be shown to satisfy the longitudinal sum rule based on the WTI, an explicit proof is somewhat more difficult. For free particle dispersion there is no difficulty. The complication is due to the fact that the bare vertex $\boldsymbol{\gamma}(\widetilde{K}, K)$ can be written down only in the small $\mathbf{q}$ limit in a periodic potential. Here one imposes the semi-classical approximation, appropriate to $\mathbf{q} \to 0$, so that the bare vertices are given by

$$\gamma_i^{\mu}(\widetilde{K}, K) = \left(1, \frac{\partial \xi_{\mathbf{k}+\mathbf{q}/2, i}}{\partial \mathbf{k}}\right). \tag{21}$$

For example, in the case where $C_1 = C_2 = 0$, and in the limit that $\mathbf{q} \to 0$, the current-current correlation function becomes

$$\overset{\leftrightarrow}{P}(Q) = 2 \sum_{K} G(\widetilde{K}) \Bigg\{ \frac{\partial \xi_{\mathbf{k}+\mathbf{q}/2}}{\partial \mathbf{k}}$$
$$+ \Delta_1^2 G_{1,1}(-K) G_{1,1}(-\widetilde{K}) \left[ -\frac{\partial \xi_{\mathbf{k}+\mathbf{q}/2, 2}}{\partial \mathbf{k}} + \Delta_2^2 G_{0,4}(K) G_{0,4}(\widetilde{K}) \frac{\partial \xi_{\mathbf{k}+\mathbf{q}/2, 4}}{\partial \mathbf{k}} \right]$$
$$+ \Delta_2^2 G_{1,2}(-K) G_{1,2}(-\widetilde{K}) \left[ -\frac{\partial \xi_{\mathbf{k}+\mathbf{q}/2, 1}}{\partial \mathbf{k}} + \Delta_1^2 G_{0,3}(K) G_{0,3}(\widetilde{K}) \frac{\partial \xi_{\mathbf{k}+\mathbf{q}/2, 3}}{\partial \mathbf{k}} \right] \Bigg\} G(K) \frac{\partial \xi_{\mathbf{k}+\mathbf{q}/2}}{\partial \mathbf{k}}, \tag{22}$$

which is in agreement with the form of $\overset{\leftrightarrow}{P}(0)$ obtained in Eq. (3) of the main text by imposing the absence of a Meissner effect in the normal phase. The longitudinal sum rule in this particular limit reduces to

$$\int \frac{d\omega}{\pi} \left( -\frac{\text{Im}\mathbf{q} \cdot \overset{\leftrightarrow}{P}(Q) \cdot \mathbf{q}}{\omega} \right) = 2 \sum_{\mathbf{k}} n_{\mathbf{k}}(\xi_{\mathbf{k}+\mathbf{q}} + \xi_{\mathbf{k}-\mathbf{q}} - 2\xi_{\mathbf{k}}), \tag{23}$$

which is in agreement with Appendix A of Tremblay's work[3].

## III. SPIN RESPONSE FUNCTIONS

In this section we extend our results to include spin response functions. Let $\sigma = \uparrow\downarrow$ denote spin indices with $\bar{\sigma}$ the opposite of $\sigma$. The density-density and current-current correlation functions are response functions related by the U(1)$_{\text{EM}}$ gauge symmetry and the associated Ward-Takahashi identity. The analogous spin response functions are related by the U(1)$_z$ gauge symmetry and the associated Ward-Takahashi identity.

The familiar U(1)$_{\text{EM}}$ gauge theory is based on the four-vector potential $A^\mu = (\phi, \mathbf{A})$. For the U(1)$_z$ gauge theory the external vector field is $A^\mu = (B_z, \mathbf{m})$, where $\mathbf{m}$ is the magnetization. The associated Hamiltonian describes a generalized spin-magnetic field interaction, and the Noether current for the global U(1)$_z$ symmetry is a magnetization current. An important difference between these two symmetries is that below $T_C$ the U(1)$_{\text{EM}}$ gauge symmetry is spontaneously broken. Thus to restore gauge invariance collective mode effects must be incorporated. On the other hand the U(1)$_z$ gauge symmetry is not spontaneously broken and therefore does not require any collective physics. Since we are only considering the response functions above $T_C$ this difference is not central to our discussion.

The bare spin vertex is denoted by $\gamma^\mu_{S_\sigma}(\widetilde{K}, K)$, where $S_\sigma = \pm 1$ and $S_{\bar{\sigma}} = -S_\sigma$ The bare and full Ward-Takahashi identities for the spin vertices are discussed in the main text [see Eq. (12) of the main text].

The full spin response function $P_S^{\mu\nu}(Q)$ is defined by

$$P_S^{\mu\nu}(Q) = \sum_\sigma \sum_K G(\widetilde{K}) \Gamma^\mu_{S_\sigma}(\widetilde{K}, K) G(K) \gamma^\nu_{S_\sigma}(K, \widetilde{K}). \tag{24}$$

Following the main text, given the form of the self energy the WTI can be used to extract the full vertex, for both the charge and spin response functions. Explicitly, the exact spin response functions are

$$\begin{aligned}
P_S^{\mu\nu}(Q) = \sum_\sigma \sum_K G(\widetilde{K}) \bigg\{ & \gamma^\mu_{S_\sigma}(\widetilde{K}, K) \\
& + \Delta_1^2 G_{1,1}(-K) \left[ \gamma^\mu_{1,S_\sigma}(-K, -\widetilde{K}) + \Delta_2^2 G_{0,4}(\widetilde{K}) \gamma^\mu_{4,S_\sigma}(\widetilde{K}, K) G_{0,4}(K) \right] G_{1,1}(-\widetilde{K}) \\
& + \Delta_2^2 G_{1,2}(-K) \left[ \gamma^\mu_{2,S_\sigma}(-K, -\widetilde{K}) + \Delta_1^2 G_{0,3}(\widetilde{K}) \gamma^\mu_{3,S_\sigma}(\widetilde{K}, K) G_{0,3}(K) \right] G_{1,2}(-\widetilde{K}) \\
& + C_1^2 G_{1,3}(\widetilde{K}) \left[ \gamma^\mu_{3,S_\sigma}(\widetilde{K}, K) + \Delta_1^2 G_{0,2}(-K) \gamma^\mu_{2,S_\sigma}(-K, -\widetilde{K}) G_{0,2}(-\widetilde{K}) \right] G_{1,3}(K) \\
& + C_2^2 G_{1,4}(\widetilde{K}) \left[ \gamma^\mu_{4,S_\sigma}(\widetilde{K}, K) + \Delta_2^2 G_{0,1}(-K) \gamma^\mu_{1,S_\sigma}(-K, -\widetilde{K}) G_{0,1}(-\widetilde{K}) \right] G_{1,4}(K) \\
& + \Delta_1 \Delta_2 C_1 G_{1,3}(K) G_{0,2}(-K) \gamma^\mu_{2,S_\sigma}(-K, -\widetilde{K}) G_{0,2}(-\widetilde{K}) \\
& - \Delta_1 \Delta_2 C_1 G_{0,2}(-\widetilde{K}) G_{1,3}(\widetilde{K}) \left[ \gamma^\mu_{3,S_\sigma}(\widetilde{K}, K) + \Delta_1^2 G_{0,2}(-K) \gamma^\mu_{2,S_\sigma}(-K, -\widetilde{K}) G_{0,2}(-\widetilde{K}) \right] G_{1,3}(K) \\
& + \Delta_1 \Delta_2 C_1 G_{1,2}(-K) \left[ \gamma^\mu_{2,S_\sigma}(-K, -\widetilde{K}) + \Delta_1^2 G_{0,3}(\widetilde{K}) \gamma^\mu_{3,S_\sigma}(\widetilde{K}, K) G_{0,3}(K) \right] G_{1,2}(-\widetilde{K}) G_{0,3}(\widetilde{K}) \\
& - \Delta_1 \Delta_2 C_1 G_{0,3}(\widetilde{K}) \gamma^\mu_{3,S_\sigma}(\widetilde{K}, K) G_{0,3}(K) G_{1,2}(-K) \\
& + \Delta_1 \Delta_2 C_2 G_{1,4}(K) G_{0,1}(-K) \gamma^\mu_{1,S_\sigma}(-K, -\widetilde{K}) G_{0,1}(-\widetilde{K}) \\
& - \Delta_1 \Delta_2 C_2 G_{0,1}(-\widetilde{K}) G_{1,4}(\widetilde{K}) \left[ \gamma^\mu_{4,S_\sigma}(\widetilde{K}, K) + \Delta_2^2 G_{0,1}(-K) \gamma^\mu_{1,S_\sigma}(-K, -\widetilde{K}) G_{0,1}(-\widetilde{K}) \right] G_{1,4}(K) \\
& + \Delta_1 \Delta_2 C_2 G_{1,1}(-K) \left[ \gamma^\mu_{1,S_\sigma}(-K, -\widetilde{K}) + \Delta_2^2 G_{0,4}(\widetilde{K}) \gamma^\mu_{4,S_\sigma}(\widetilde{K}, K) G_{0,4}(K) \right] G_{1,1}(-\widetilde{K}) G_{0,4}(\widetilde{K}) \\
& - \Delta_1 \Delta_2 C_2 G_{0,4}(\widetilde{K}) \gamma^\mu_{4,S_\sigma}(\widetilde{K}, K) G_{0,4}(K) G_{1,1}(-K) \bigg\} G(K) \gamma^\nu_{S_\sigma}(K, \widetilde{K}).
\end{aligned} \tag{25}$$

The proof of the $f$-sum rule in section (II. A) can be performed for the spin response function $P_S^{00}(Q)$ in an analogous manner, by using the WTI for the spin vertices. In this case we obtain

$$\int \frac{d\omega}{\pi}(-\omega \text{Im} P_S^{00}(Q)) = 2 \sum_{\mathbf{k}} n_{\mathbf{k}}(\xi_{\mathbf{k+q}} + \xi_{\mathbf{k-q}} - 2\xi_{\mathbf{k}}), \tag{26}$$

where $n_{\mathbf{k}} = T \sum_{i\omega} G(K)$ and the factor of two arises from summation over pseudo spin indices.

## IV. EXTENSION BELOW THE TRANSITION TEMPERATURE

This paper deals exclusively with the normal state. However, it is often convenient to compare and contrast the behavior above and below $T_C$. Given the mean field form of the normal state self energy $\Sigma_{pg}$ it is natural to follow earlier work of our own group[4] and of Yang, Rice and Zhang (YRZ)[5] and address the broken symmetry state by taking the full self energy, $\Sigma$, to consist of two terms: $\Sigma = \Sigma_{pg} + \Sigma_{sc}$. Here $\Sigma_{sc}$ is another BCS mean-field like self energy corresponding to the presumed form of the condensate. In this way, there are two different gap parameters: $\Delta_{sc}$ and $\Delta_{pg}$.

There are several important distinctions in the way in which these two self energy terms enter into response functions. As suggested in our earlier work[4,6] for the $d$-wave pseudogap and for the YRZ case[7], the self energy involving $\Delta_{pg}$ should also contain a damping term representing the fact that the non-condensed pairs are not infinitely long lived. This gives rise to the arcs (or spread out nodes) in the case of a $d$-wave pseudogap.

There are even more important features associated with the sign of the contributions from $\Delta_{sc}$ and $\Delta_{pg}$ in various response functions, which must guarantee that the superfluid density depends only on the condensate $\Delta_{sc}$; these sign changes appear in the charge density and current density response functions[4,6]. In a related fashion, the density-density correlation function must include collective mode effects below the transition, in order to be consistent with sum rules.

However, there is no such sign change of $\Sigma_{pg}$ relative to $\Sigma_{sc}$ in the spin response, which is not associated with a Meissner effect. Thus in the case of a simple $d$-wave pseudogap and in the YRZ case as well, one does not expect there to be a significant difference in the spin correlation functions above and below $T_C$.

By contrast, in the Amperean pairing model[8], it is argued that at $T_C$ the finite wave vector pairing gap converts to a conventional $d$-wave superconducting phase. This gives rise to a dramatic change in the features associated with neutron scattering. Indeed, a moderate change through the transition is observed in some experiments[9], but, nevertheless, peaks at or near $(\pi, \pi)$ are still present. This is in contrast to the rather structureless behavior of the cross section which is found in the Amperean pairing theory reported in this paper.

---


\end{document}